\documentclass[aps,prd,print,groupedaddress,nofootinbib,eqsecnum,onecolumn]{revtex4}
\usepackage{eurosym}
\usepackage{amsmath}
\usepackage{amsthm}
\usepackage{amssymb}
\usepackage{graphicx}
\usepackage{slashed}
\usepackage{hyperref}
\usepackage{epstopdf}

\theoremstyle{plain}

\theoremstyle{definition}

\begin{document}

\author{Michael Rios}
\email{mrios@dyonicatech.com}
\affiliation{Dyonica ICMQG,\\
Los Angeles, CA, 90032, USA}
\author{Alessio Marrani}
\email{jazzphyzz@gmail.com}
\affiliation{Museo Storico della Fisica e Centro Studi e Ricerche Enrico Fermi, \\
Via Panisperna 89A, I-00184 Roma, Italy}
\author{David Chester}
\email{dchester@ucla.edu}
\affiliation{Department of Physics and Astronomy,\\
UCLA, Los Angeles, CA, 90095, USA}
\title{Geometry of exceptional super Yang-Mills theories}

\begin{abstract}
Some time ago, Bars found $D=11+3$ supersymmetry and Sezgin proposed super Yang-Mills
theory (SYM) in $D=11+3$. Using the ``\textit{magic star}''
projection of $\mathfrak{e}_{8(-24)}$, we show that the geometric structure of
SYM's in $12+4$ and $11+3$ space-time dimensions descends to the
affine symmetry of the space $AdS_{4}\otimes S^{8}$. By reducing to transverse transformations
along maximal embeddings, the near-horizon geometries of the M2 brane 
($AdS_{4}\otimes S^{7}$) and M5 brane ($AdS_{7}\otimes S^{4}$) of M-theory are recovered.
Utilizing the recently
introduced \textit{``exceptional periodicity"} (EP) and exploiting 
the embedding of semisimple rank-3 Jordan algebras into rank-3 T-algebras of special type 
yields the spaces $AdS_{4}\otimes S^{8n}$ and $AdS_{8n-1}\otimes
S^{5}$ with reduced subspaces $AdS_{4}\otimes S^{8n-1}$ and 
$AdS_{8n-1}\otimes S^{4}$, respectively. As such, EP describes the 
near-horizon geometries of an infinite class of novel \textit{exceptional} SYM's in $(8n+3)+3$ dimensions
that generalize M-theory for $n=1$. Remarkably, the $n=3$ level hints at M2 and M21
branes as solutions of bosonic M-theory and gives support for Witten's
monstrous $AdS$/CFT construction.
\newline
\newline
$Keywords:$ Yang-Mills, exceptional periodicity, Monster CFT, M-theory,
Bosonic M-theory.
\end{abstract}

\maketitle

\vspace{7.0cm}

%\date{\today}

%\vspace{0.3cm}

%\vspace{0.3cm}

%\vspace{1.0cm}

%\vspace{4.0cm}

%\author{David Chester}
%\email{chester31124@gmail.com}
%\affiliation{
%Department of Physics and Astronomy,\\
%UCLA, Los Angeles, CA 90095-1547, USA}

% Beyond the Standard Model and M-theory
% On $D=14$ Super-Yang-Mills Theory, A-Theory, the Standard Model, and Gravity
\setcounter{page}{1}

\newpage

\tableofcontents
\addcontentsline{toc}{section}{Acknowledgements}

%\newpage

\section{Introduction}

After Witten proposed M-theory \cite{WittenM}, arguments to go beyond $D=11$
have been put forth by Vafa \cite{VafaF}, Bars \cite{BarsS,2}, Sezgin \cite{1} 
and Nishino \cite{6}. Bars found success in his model of S-theory with
two time dimensions in $D=s+t=11+2$ \cite{BarsS} and proposed $D=11+3$ 
supersymmetry with three times~\cite{2}, while Sezgin
formulated super Yang-Mills theories (SYM's) up to signature $11+3$ \cite{1}%
. The signature $11+3$ also found promise in graviGUT models \cite{Lisi,
Percacci}. Nishino pushed beyond $D=11+3$ and gave arguments for SYM's in
signature $\left( 9+m\right) +\left( 1+m\right) $, for arbitrary $m\in
\mathbb{N}\cup \left\{ 0\right\} $.

The symmetry of $11+3$ space-time, with a $64$-dimensional Majorana-Weyl
(MW) spinor, interestingly arises in a certain 5-grading of
\textquotedblleft extended Poincar\'{e} type" of $\mathfrak{e}_{8(-24)}$,
while the symmetry of $10+2$ space-time is seen in a 5-grading of
\textquotedblleft contact type" of $\mathfrak{e}_{7(-25)}$, with a $32$%
-dimensional MW spinor \cite{Santi-KTS}. Moreover, a certain 3-grading of $%
\mathfrak{e}_{6(-26)}$ contains the symmetry of $9+1$ space-time with a $16$%
-dimensional MW spinor. These are precisely the signatures studied by Bars and Sezgin
up to $11+3$ \cite{2,1}.

Through a projection of $\mathfrak{e}_{8(-24)}$ along an $\mathfrak{sl}_{3,%
\mathbb{R}}$ subalgebra (the so-called \textit{\textquotedblleft magic star"}
projection, cf. Fig. \ref{fig:MagicStar}), the hidden Jordan algebraic
structure within $\mathfrak{e}_{8(-24)}$ becomes manifest \cite{3,4bis,4,5}.
The central vertex $\mathfrak{e}_{6(-26)}$ of the projection then encodes
the reduced structure symmetry of the degree 3 exceptional Jordan
algebra (also known as Albert algebra) $\mathbf{J}_{3}^{\mathbb{O}}$, mapped
to six vertices of the star projection.

Using the aforementioned gradings of $\mathfrak{e}_{6(-26)}$, $\mathfrak{e}%
_{7(-25)}$ and $\mathfrak{e}_{8(-24)}$, as well as the magic star, a
periodic extension of the exceptional Lie algebras can be formulated. This
periodic extension, dubbed \textit{exceptional periodicity} (EP) \cite{5},
allows higher dimensional extensions of the exceptional Lie algebras, while
also permitting arbitrarily high dimensional magic star projections.
Although the resulting algebras (with generalized roots) no longer satisfy
the Jacobi relation, they do contain Lie algebraic reductive parts, as well
as a normalized cocycle, as seen in lattice vertex algebras \cite%
{5,forthcoming2}.\medskip

In this study, using some gradings of the exceptional Lie algebras and their
exceptionally periodic extensions, we show that the geometric structure of
SYM's in $11+3$ and $12+4$ space-time dimensions can be recovered from the
affine symmetry of the space $AdS_{4}\otimes S^{8}$, with the $8$-sphere
being a line in the Cayley plane $\mathbb{OP}^{2}$, the space of rank-$1$
projectors of $J_{3}^{\mathbb{O}}$ \cite{12,cattoProj}. The symmetry of $S^{8}$, $SO(9)
$ (the light cone little group of M-theory), is a maximal and symmetric
subgroup of $F_{4}$, and as such is the stabilizer of $\mathbb{OP}^{2}\simeq
F_{4}/SO(9)$ itself. A fixed point of $\mathbb{OP}^{2}$, a rank-1 idempotent
of $J_{3}^{\mathbb{O}}$, identifies one of three possible embeddings $%
SO(9)\subset F_{4}$ and spans an orthogonal direction that can serve as the
11th dimension of M-theory. By considering transverse directions along
maximal embeddings, the near-horizon geometries of the M2 brane ($%
AdS_{4}\otimes S^{7}$) and M5 brane ($AdS_{7}\otimes S^{4}$) are recovered.

Generalizing the construction to higher levels of exceptional periodicity
(parametrized by $n\in \mathbb{N}$), where Jordan algebras of degree three
are lifted to the special class of rank-3 Vinberg's T-algebras \cite{Vinberg}%
, maximal embeddings that respect the symmetry of the T-algebraic spin
factors yield the spaces $AdS_{4}\otimes S^{8n}$ and $AdS_{8n-1}\otimes S^{5}
$, with reduced subspaces $AdS_{4}\otimes S^{8n-1}$ and $AdS_{8n-1}\otimes
S^{4}$, respectively. Through exceptional periodicity, this suggests
generalizations of the M2 brane and M5 brane near-horizon geometries from
SYM's in $\left( 8n+3\right) +3$ space-time dimensions, as descending from
SYM's in $\left( 8n+4\right) +4$ space-time dimensions, resulting in an M$(8n-3)$ brane dual to the M2 brane\cite{7}.\medskip

The plan of the paper is as follows.

Within EP, we explicitly study levels $n=1$ (the trivial level,
corresponding to exceptional Lie algebras, in particular to $\mathfrak{e}%
_{8(-24)}$ in our case), $n=2$ (the first nontrivial level) and $n=3$,
respectively in Secs. \ref{n=1}, \ref{n=2} and \ref{n=3}. Interestingly, at
level $n=3$, the resulting branchings give $AdS_4 \otimes S^{23}$ and hint at M2 and M21 branes as solutions
of the (conjectured) bosonic M-theory \cite{14}; by reduction to $%
AdS_{3}\otimes S^{23}$, we recover a space that lends support for Witten's
monstrous $AdS$/CFT construction for three-dimensional gravity \cite{13}, as
the Conway group $Co_{0}$ (the symmetry of the Leech lattice \cite{16}) is
recovered from the $SO(24)$ $\mathcal{R}$-symmetry\footnote{%
For $D=27$ M-theory reduced to $D=3$, there is no light cone little group and
the expected $\mathcal{R}$-symmetry is the full $SO(24)$, the transversal
rotation group in $D=26$. This is the higher dimensional analog of what
occurs for $SO(8)$ $\mathcal{R}$-symmetry in $D=11$ M-theory reduced to $D=3$%
.} of a discretized $S^{23}$%
. On the other hand, our analysis at levels $n=2$ and $3$ can be bridged by
the observation that every K3 sigma model contains a symmetry group that is
a subgroup of $Co_{0}$ \cite{Mathieu}.

Section \ref{Higher-SYMs} therein presents evidence for the existence of higher
dimensional $\mathcal{N}=(1,0)$ SYM's with $1$, $2$, $3$ or $4$ timelike
dimensions, named \textit{exceptional SYM's}, stemming from similar
constructions given by Bars~\cite{2}, Sezgin~\cite{1,1bis}, and Nishino~\cite{6}.

Such higher dimensional SYM's can be defined at every level of EP for $%
\mathfrak{e}_{8(-24)}$, whose generic $n$th level is considered in Section \ref%
{n}. The resulting ``EP/SYM correspondence" (which
will be investigated further in suggests a spectral formulation of M-theory
from the class of Vinberg's special cubic T-algebras \cite{Vinberg}. This
generalizes the structure of matrix theory \cite{matrixM} in $D=10+1$ to a
more general class of nonassociative matrix operator algebras that
periodically exhibit nonassociative geometry, up to infinite
dimensions. This is as Connes \cite{ncg} has done for noncommutative
geometry from noncommutative $C^{\ast }$-algebras, which falls under the
more general \textit{motivic} program of Grothendieck \cite{motive}.

\section{\label{n=1}$\mathfrak{e}_{8(-24)}$}

\subsection{$\mathcal{N}=(1,0)$ SYM in $11+3$ dimensions}

\begin{figure}[t]
\centering
\includegraphics[width=0.60\textwidth]{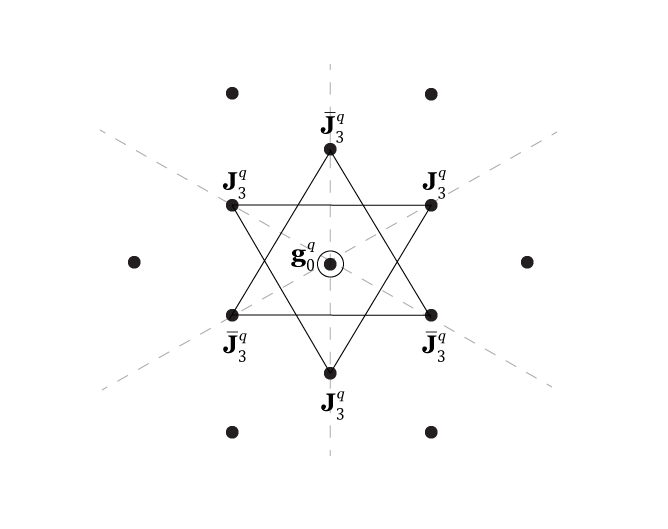}
\caption{The \textit{\textquotedblleft Magic Star"} of finite-dimensional
exceptional Lie algebras \protect\cite{3}. $\mathbf{J}_{3}^{q}$ denotes the
Jordan algebra of $3\times 3$ Hermitian matrices over the division algebras $%
\mathbb{A}=\mathbb{R}$, $\mathbb{C}$, $\mathbb{H}$ and $\mathbb{O}$ for $%
q=\dim _{\mathbb{R}}\mathbb{A}=1,2,4,8$ respectively, whereas $\mathbf{g}_{%
\mathbf{0}}^{q}$ denotes its reduced structure Lie algebra. In the case of
the minimally noncompact, real form $\mathfrak{e}_{8(-24)}$ of $\mathfrak{e}%
_{8}$ under consideration, $q=8$ and $\mathbf{g}_{0}^{8}=\mathfrak{e}%
_{6(-26)}$. In the text, we use $\mathbf{J}_{3}^{\mathbb{O}}\equiv \mathbf{J}%
_{3}^{8}$.}
\label{fig:MagicStar}
\end{figure}

The Cayley plane $\mathbb{OP}^{2}$ is the projective space of all rank-$1$
projectors of the exceptional Jordan algebra $\mathbf{J}_{3}^{\mathbb{O}}$;
by stabilizing a point of the Cayley plane, the affine symmetry $E_{6(-26)}$
is reduced to $SO(9,1)$ \cite{12}. This $SO(9,1)$ subgroup acts on a line
(an $S^{8}$) of the Cayley plane as affine transformations. It is also the
symmetry of the $10$-dimensional spin factor from the Peirce decomposition
(cf., e.g., \cite{Jacobson,Schafer})%
\begin{equation}
\mathbf{J}_{3}^{\mathbb{O}}=\mathbf{10}\oplus \mathbf{16}\oplus \mathbf{1},%
\label{pd}
\end{equation}%
where the fixed point (primitive idempotent of $\mathbf{J}_{3}^{\mathbb{O}}$)
corresponds to the $\mathbf{1}$ in the rhs of (\ref{pd}), and it serves
as a point at infinity for the eight-dimensional transverse space. The affine
symmetry and MW semispinor $\mathbf{16}$ occur in the 3-grading\footnote{%
This is a remarkable 3-grading structure, related to a Jordan pair that is
not a pair of Jordan algebras, but rather a Jordan triple system; cf., e.g.,
\cite{Truini-E6}.} of $\mathfrak{e}_{6(-26)}$,
\begin{equation}
\mathfrak{e}_{6(-26)}=\mathbf{16}_{-3}^{\prime }\oplus (\mathfrak{so}%
_{9,1}\oplus \mathbb{R})_{0}\oplus \mathbf{16}_{+3},  \label{aa}
\end{equation}%
where $\mathbf{16}^{\prime }$ is the conjugate MW semispinor in $9+1$
space-time dimensions. (\ref{aa}) can be interpreted as the maximal
symmetric embedding%
\begin{equation}
\mathfrak{str}_{0}\left( \mathbf{J}_{3}^{\mathbb{O}}\right) \supset
\mathfrak{str}_{0}\left( \mathbb{R\oplus }\mathbf{J}_{2}^{\mathbb{O}}\right)
,
\end{equation}%
which is a consequence of the maximal Jordan-algebraic embedding,%
\begin{equation}
\mathbf{J}_{3}^{\mathbb{O}}\supset \mathbb{R\oplus }\mathbf{J}_{2}^{\mathbb{O%
}},  \label{JAE}
\end{equation}%
where $\mathbf{J}_{2}^{\mathbb{O}}\simeq \mathbf{\Gamma }_{9,1}$ (cf., e.g.,
(5.2) of \cite{G-Lects}) is a $10$-dimensional Lorentzian spin factor \cite%
{huertaSYM, smolinMatrix} and $\mathbb{R}$ is the span of the fixed idempotent\footnote{%
The $\mathbb{R}$ idempotent direction transforms spatially under $\mathfrak{%
so}_{10,1}\subset \mathfrak{so}_{10,2}\subset \mathfrak{e}_{7(-25)}$, and
can serve as an 11th dimension of M-theory.}.

The magic star of $\mathfrak{e}_{8(-24)}$ (cf.
Fig. \ref{fig:MagicStar} and its caption) is a projection along an $%
\mathfrak{sl}_{3,\mathbb{R}}$ subalgebra, with $\mathfrak{e}_{6(-26)}$
projected to a central vertex \cite{3,4bis,4,5}. In $\mathfrak{e}_{8(-24)}$,
the analog of $\mathfrak{so}_{9,1}$ is $\mathfrak{so}_{12,4}$; indeed, the
following maximal symmetric embedding holds,%
\begin{equation}
\mathfrak{e}_{8(-24)}\supset \mathfrak{so}_{12,4},  \label{a}
\end{equation}%
interpreted as%
\begin{equation}
\mathfrak{qconf}\left( \mathbf{J}_{3}^{\mathbb{O}}\right) \supset \mathfrak{%
qconf}\left( \mathbb{R\oplus }\mathbf{J}_{2}^{\mathbb{O}}\right) ,
\end{equation}%
which is still a consequence of (\ref{JAE}); $\mathfrak{qconf}$ here denotes
the quasiconformal symmetry of the corresponding rank-3 Jordan algebra \cite%
{quasiconformal}. The branching corresponding to (\ref{a}) gives naturally
rise to the following 5-grading%
\begin{equation}
\mathfrak{e}_{8(-24)}=\mathfrak{so}_{12,4}\oplus \mathbf{128}=\mathbf{14}%
_{-2}\oplus \mathbf{64}_{-1}^{\prime }\oplus (\mathfrak{so}_{11,3}\oplus
\mathbb{R})_{0}\oplus \mathbf{64}_{+1}\oplus \mathbf{14}_{+2},  \label{a-3}
\end{equation}%
where $\mathbf{64}$ and $\mathbf{64}^{\prime }$ are the MW semispinor and
its conjugate in $D=s+t=11+3$ space-time dimensions. Part of (\ref{a-3}) has
already appeared in \cite{Lisi, Percacci}. As mentioned above,
mathematically, it identifies (the minimally noncompact, real form of) a
Kantor triple system, of \textquotedblleft extended Poincar\'{e} type", over
$\mathfrak{e}_{8}$ \cite{Santi-KTS}. Moreover, we recall that $\mathfrak{so}%
_{11,3}$, occurring in the 0-graded reductive component of the 5-grading (%
\ref{a-3}), is the space-time, purely bosonic, symmetry Lie algebra of the $%
\mathcal{N}=(1,0)$ SYM studied by Sezgin, Bars and Nishino in $D=11+3$
space-time dimensions \cite{1,2,6} (see also Section \ref{Higher-SYMs} below).

\subsection{M2 brane in $10+1$ dimensions}

At the Lie group level, a maximal symmetric subgroup of $SO(12,4)$ is
\begin{equation}
SO(12,4)\supset SO(3,3)\times SO(9,1).
\end{equation}%
Recalling the coset
\begin{equation}
AdS_{4}=O(3,2)/O(3,1),
\end{equation}%
$SO(3,2)$ acts on $AdS_{4}$ as isometries, while $SO(3,3)$ acts via affine
transformations. Hence, $SO(12,4)$ includes maximally (and symmetrically)
the affine transformations of $AdS_{4}\otimes S^{8}$.

Reducing to transversal rotations, the isometry group $SO(9)$ of $%
S^{8}\simeq \mathbb{OP}^{1}$ (which can be regarded as a line in $\mathbb{OP}%
^{2}$) breaks to $SO(8)$ acting on $S^{7}$. This corresponds to the two-step
chain of maximal symmetric embeddings into $SO(12,4)$,
\begin{equation}
SO(12,4)\supset SO(11,3)\supset SO(3,3)\times SO(8),
\end{equation}%
breaking down to isometries of $AdS_{4}$ by further maximal and symmetric
embedding yields%
\begin{equation}
SO(12,4)\supset SO(11,3)\supset SO(3,3)\times SO(8)\supset SO(3,2)\times
SO(8).
\end{equation}%
Alternatively, the following embedding also holds,
\begin{equation}
SO(12,4)\supset SO(3,2)\times SO(8)\times SO(1,2),
\end{equation}%
where $SO(3,2)\times SO(8)$ acts as isometries of $AdS_{4}\otimes S^{7}$,
the near-horizon geometry of the M2 brane [whose world volume symmetry is
described by $SO(1,2)$], as a solution of $10+1$ M-theory (or of its
low-energy limit, $\mathcal{N}=1$ eleven-dimensional supergravity); see e.g.
\cite{9,10,11}.

\subsection{M5 brane in $10+1$ dimensions}

Another maximal symmetric subgroup of $SO(12,4)$ is
\begin{equation}
SO(12,4)\supset SO(6,3)\times SO(6,1).
\end{equation}%
Recalling the coset
\begin{equation}
AdS_{7}=O(6,2)/O(6,1),
\end{equation}%
$SO(6,2)$ acts on $AdS_{7}$ as isometries, and $SO(6,3)$ acts via affine
transformations. Hence, $SO(12,4)$ includes maximally (and symmetrically)
the affine transformations of $AdS_{7}\otimes S^{5}$.

Considering the reduction $S^{5}\rightarrow S^{4}$, where $S^{4}\simeq
\mathbb{HP}^{1}$ can be conceived as a line in $\mathbb{HP}^{2}$, the
isometry group $SO(6)$ of $S^{5}$ breaks to $SO(5)$ on $S^{4}$. Thus, the
reduction $\mathbb{O}\rightarrow\mathbb{H}$ reduces $\mathbb{OP}^1\simeq
S^8\rightarrow \mathbb{HP}^1\simeq S^4$ with the point at infinity given by a
fixed primitive idempotent of $\mathbf{J}_{3}^{\mathbb{H}}$. This
corresponds to the following chain of maximal symmetric embeddings into $%
SO(12,4)$,
\begin{equation}
SO(12,4)\supset SO(11,3)\supset SO(6,3)\times SO(5),  \label{ee}
\end{equation}%
breaking down to isometries of $AdS_{7}$ by further maximal and symmetric
embedding yields%
\begin{equation}
SO(12,4)\supset SO(11,3)\supset SO(6,3)\times SO(5)\supset SO(6,2)\times
SO(5).
\end{equation}%
Alternatively, the following embedding also holds,
\begin{equation}
SO(12,4)\supset SO(6,2)\times SO(5)\times SO(1,2),  \label{M5}
\end{equation}%
where $SO(6,2)\times SO(5)$ acts as isometries of $AdS_{7}\otimes S^{4}$,
the near-horizon geometry of the M5 brane, Hodge dual to the M2 brane [whose
world volume symmetry $SO(1,2)$ still occurs as a commuting factor] in $10+1$
space-time dimensions (see e.g. \cite{9,10,11}).

Note also that $SO(12,4)$ contains the isometries of $AdS_{7}$ and of the
M5 brane worldvolume, times a dilatational factor,%
\begin{equation}
SO(12,4)\supset SO(6,2)\times SO(5,1)\times SO(1,1)\simeq SO^{\ast
}(8)\times SU^{\ast }(4)\times SO(1,1),
\end{equation}%
or, equivalently, through two different maximal and symmetric embeddings,
the isometries of $AdS_{7}$ times the conformal symmetry of the M5 brane
worldvolume, or, respectively, the conformal symmetry of $AdS_{7}$ times the
isometries of the M5 brane worldvolume,%
\begin{eqnarray}
SO(12,4) &\supset &SO(6,2)\times SO(6,2)\simeq SO^{\ast }(8)\times SO^{\ast
}(8), \\
SO(12,4) &\supset &SO(7,3)\times SO(5,1)\simeq SO(7,3)\times SU^{\ast }(4).
\end{eqnarray}

\section{\label{EP}Generalizations within exceptional periodicity}

\begin{figure}[tp]
\centering
\includegraphics[width=0.60\textwidth]{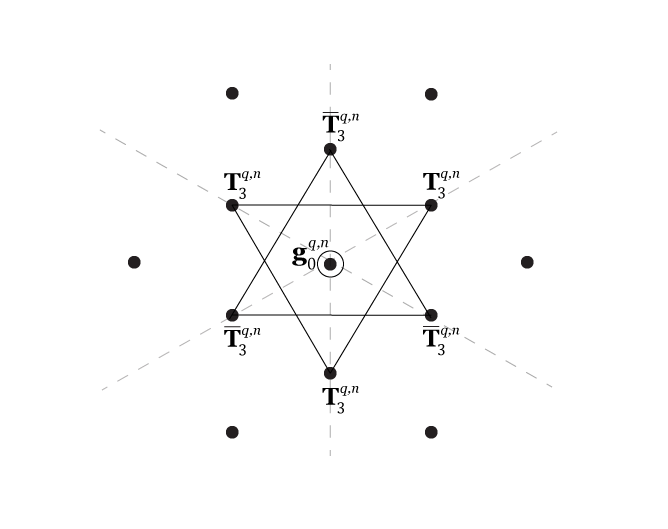}
\caption{The \textit{\textquotedblleft magic star"} of EP \protect\cite{5}. $%
\mathbf{T}_{3}^{q,n}$ denotes a T-algebra of rank-3 and of \textit{special
type} \protect\cite{Vinberg}, parametrized by $q=\dim _{\mathbb{R}}\mathbb{A}%
=1,2,4,8$ for $\mathbb{A}=\mathbb{R},\mathbb{C},\mathbb{H},\mathbb{O}$
respectively, and $n\in \mathbb{N}$ \protect\cite{Vinberg,Marrani-Group32}.
In the case of the EP for $\mathfrak{e}_{8(-24)}$, as under discussion, $q=8$%
.}
\label{fig:MagicStarT}
\end{figure}

\subsection{\label{Higher-SYMs}Exceptional SYM's}

As mentioned in the introduction, various efforts have been made over the
years to define consistent supersymmetric theories in $D>11$ space-time
dimensions. For instance, Nishino constructed an $\mathcal{N}=(1,0)$ SYM
\cite{1-N} as well as an $\mathcal{N}=1$ \cite{2-N} and $\mathcal{N}=2$ \cite%
{3-N} supergravity theory in $10+2$ dimensions, with the two timelike
dimensions being motivated by the development of F-theory \cite{VafaF, 5-N,
6-N}, Bars' S-theory \cite{BarsS}, or other two-times theories \cite{8-N}.
Nonminimal, chiral, $\mathcal{N}=(2,0)$ SYM in $10+2$ was considered by
Sezgin (cf., e.g., \cite{1bis}). More recently, Castellani proposed an $%
\mathcal{N}=1$ supergravity in $10+2$ with locally supersymmetric $SO(10,2)$%
-invariant action \cite{Castellani}. Switching to three timelike dimensions, Bars introduced $11+3$ supersymmetry~\cite{2} and
Sezgin constructed an $\mathcal{N}=(1,0)$ SYM in $11+3$~\cite{1},
obtaining a space-time superalgebra with 64 real supercharges containing
the type IIB Poincar\'{e} superalgebra in $9+1$ as well as $\mathcal{N}=1$
Poincar\'{e} superalgebra in $10+1$. However, Sezgin also found an
obstruction for extending global supersymmetry beyond $11+3$, later overcome
by Nishino~\cite{6}, who formulated $\mathcal{N}=(1,0)$ SYM's in any
signature $\left( 9+m\right) +\left( 1+m\right) $, for arbitrary $m\in
\mathbb{N}\cup \left\{ 0\right\} $, with four general classes of SYM's given
by the fact that $m=4\mathbf{n}$, $\mathbf{n}\in \mathbb{N}\cup \left\{
0\right\} $.

Essentially, Bars~\cite{2}, Sezgin \cite{1}, and then Nishino \cite{6} considered the
symmetries of the $\gamma$-matrices \cite{12-N}, in particular, signatures $%
D=s+t$, all with $s-t=8$. Disregarding the Euclidean case ($t=0$), four $%
\mathbf{n}$-parametrized classes of chiral, minimal $\mathcal{N}=(1,0)$
superalgebras can be consistently defined \cite{1,6},

\begin{enumerate}
\item $D=\left( 9+4\mathbf{n}\right) +$ $\left( 1+4\mathbf{n}\right) $:%
\begin{equation}
\left\{ Q_{\alpha },Q_{\beta }\right\} =\left( \gamma ^{\mu }\right)
_{\alpha \beta }P_{\mu }+\left( \gamma ^{\mu _{1}...\mu _{5}}\right)
_{\alpha \beta }Z_{\mu _{1}....\mu _{5}}+...+\left( \gamma ^{\mu _{1}...\mu
_{5+4\mathbf{n}}}\right) _{\alpha \beta }Z_{\mu _{1}....\mu _{5+4\mathbf{n}%
}}.  \label{1-t}
\end{equation}

\item $D=\left( 10+4\mathbf{n}\right) +$ $\left( 2+4\mathbf{n}\right) $:%
\begin{equation}
\left\{ Q_{\alpha },Q_{\beta }\right\} =\left( \gamma ^{\mu \nu }\right)
_{\alpha \beta }Z_{\mu \nu }+\left( \gamma ^{\mu _{1}...\mu _{6}}\right)
_{\alpha \beta }Z_{\mu _{1}....\mu _{6}}+...+\left( \gamma ^{\mu _{1}...\mu
_{6+4\mathbf{n}}}\right) _{\alpha \beta }Z_{\mu _{1}....\mu _{6+4\mathbf{n}%
}}.  \label{2-t}
\end{equation}

\item $D=\left( 11+4\mathbf{n}\right) +$ $\left( 3+4\mathbf{n}\right) $:%
\begin{equation}
\left\{ Q_{\alpha },Q_{\beta }\right\} =\left( \gamma ^{\mu \nu \rho
}\right) _{\alpha \beta }Z_{\mu \nu \rho }+\left( \gamma ^{\mu _{1}...\mu
_{7}}\right) _{\alpha \beta }Z_{\mu _{1}....\mu _{7}}+...+\left( \gamma
^{\mu _{1}...\mu _{7+4\mathbf{n}}}\right) _{\alpha \beta }Z_{\mu _{1}....\mu
_{7+4\mathbf{n}}}.  \label{3-t}
\end{equation}

\item $D=\left( 12+4\mathbf{n}\right) +$ $\left( 4+4\mathbf{n}\right) $:%
\begin{equation}
\left\{ Q_{\alpha },Q_{\beta }\right\} =\eta _{\alpha \beta }Z+\left( \gamma
^{\mu _{1}...\mu _{4}}\right) _{\alpha \beta }Z_{\mu _{1}...\mu _{4}}+\left(
\gamma ^{\mu _{1}...\mu _{8}}\right) _{\alpha \beta }Z_{\mu _{1}....\mu
_{8}}+...+\left( \gamma ^{\mu _{1}...\mu _{8+4\mathbf{n}}}\right) _{\alpha
\beta }Z_{\mu _{1}....\mu _{8+4\mathbf{n}}}.  \label{4-t}
\end{equation}
\end{enumerate}

In such space-time superalgebras, $Z_{\mu _{1}...\mu _{p}}$ are the bosonic
p-form generators. Actually, in (\ref{1-t})-(\ref{4-t}) the $\gamma $%
-matrices are all chirally projected: $\gamma ^{\mu _{1}...\mu _{p}}\equiv $
$\gamma ^{\mu _{1}...\mu _{p}}C^{-1}$, with $C$ denoting the charge
conjugation matrix [the $\eta $-matrix occurring in (\ref{4-t}) is the
chiral projection of $C$ itself]. The maximal rank $\gamma $-matrices in the
rhs's of (\ref{1-t})-(\ref{4-t}) have a definite duality property, and
hence the corresponding bosonic generator is taken to be self-dual. Thus,
the rhs's of (\ref{1-t})-(\ref{4-t}) span the full symmetric space of
relevant dimension, and all the $\gamma $-matrices surviving the chiral
projection occur therein. It is here worth pointing out that, among the
above chiral superalgebras, only the one in $D=\left( 9+4\mathbf{n}\right) +$
$\left( 1+4\mathbf{n}\right) $ space-time dimensions (\ref{1-t}) is a
\textit{proper} Poincar\'{e} superalgebra, containing the momentum operator $%
P_{\mu }$ in the rhs.

We should also recall that the symmetry property of the $\gamma $-matrices
repeats itself every eight dimensions in space-time \cite{12-N}, while the
chirality (dottedness) of the spinors alternates every two dimensions. In
particular, the properties of spinors are defined by two parameters: $D=s+t$
mod($8$) and $\rho =s-t$ mod($8$) [the mod($8$) periodicity being the Bott
periodicity; see e.g.~\cite{SpinorAlgebras}, and references therein]. Thus, the
fact that all superalgebras (\ref{1-t})-(\ref{4-t}) are characterized by a
MW (semi)spinor generator can be traced back that they all have $\rho =8=0$
mod($8$).

The simple but crucial observation is that, by switching $4\mathbf{n}$
timelike dimensions into $4\mathbf{n}$ spacelike dimensions in the above
space-time signatures, one keeps $D$ and $\rho $ unchanged, and so
all the properties of spinors and $\gamma $-matrices relevant for the
definition of the above super-Poincar\'{e} algebras are left unchanged. 
Thus, one can check that the proofs given in \cite{6} (see also \cite{1})
for the existence of SYM's with superalgebras (\ref{1-t})-(\ref{2-t}) can be
successfully repeated to show the \textit{existence of SYM's} based on
global minimal, chiral $(1,0)$ superalgebras in the following space-time
signatures,

\begin{enumerate}
\item $D=\left( 9+8\mathbf{n}\right) +$ $1$:%
\begin{equation}
\left\{ Q_{\alpha },Q_{\beta }\right\} =\left( \gamma ^{\mu }\right)
_{\alpha \beta }P_{\mu }+\left( \gamma ^{\mu _{1}...\mu _{5}}\right)
_{\alpha \beta }Z_{\mu _{1}....\mu _{5}}+...+\left( \gamma ^{\mu _{1}...\mu
_{5+4\mathbf{n}}}\right) _{\alpha \beta }Z_{\mu _{1}....\mu _{5+4\mathbf{n}%
}}.\label{1-t-new}
\end{equation}

\item $D=\left( 10+8\mathbf{n}\right) +$ $2$:%
\begin{equation}
\left\{ Q_{\alpha },Q_{\beta }\right\} =\left( \gamma ^{\mu \nu }\right)
_{\alpha \beta }Z_{\mu \nu }+\left( \gamma ^{\mu _{1}...\mu _{6}}\right)
_{\alpha \beta }Z_{\mu _{1}....\mu _{6}}+...+\left( \gamma ^{\mu _{1}...\mu
_{6+4\mathbf{n}}}\right) _{\alpha \beta }Z_{\mu _{1}....\mu _{6+4\mathbf{n}%
}}.\label{2-t-new}
\end{equation}

\item $D=\left( 11+8\mathbf{n}\right) +$ $3$:%
\begin{equation}
\left\{ Q_{\alpha },Q_{\beta }\right\} =\left( \gamma ^{\mu \nu \rho
}\right) _{\alpha \beta }Z_{\mu \nu \rho }+\left( \gamma ^{\mu _{1}...\mu
_{7}}\right) _{\alpha \beta }Z_{\mu _{1}....\mu _{7}}+...+\left( \gamma
^{\mu _{1}...\mu _{7+4\mathbf{n}}}\right) _{\alpha \beta }Z_{\mu _{1}....\mu
_{7+4\mathbf{n}}}.\label{3-t-new}
\end{equation}

\item $D=\left( 12+8\mathbf{n}\right) +$ $4$:%
\begin{equation}
\left\{ Q_{\alpha },Q_{\beta }\right\} =\eta _{\alpha \beta }Z+\left( \gamma
^{\mu _{1}...\mu _{4}}\right) _{\alpha \beta }Z_{\mu _{1}...\mu _{4}}+\left(
\gamma ^{\mu _{1}...\mu _{8}}\right) _{\alpha \beta }Z_{\mu _{1}....\mu
_{8}}+...+\left( \gamma ^{\mu _{1}...\mu _{8+4\mathbf{n}}}\right) _{\alpha
\beta }Z_{\mu _{1}....\mu _{8+4\mathbf{n}}}.\label{4-t-new}
\end{equation}
\end{enumerate}

The SYM's whose space-time (super-Poincar\'{e}) superalgebra is given by (\ref%
{1-t-new})-(\ref{4-t-new}) are named exceptional SYM's
henceforth. Note that $\mathbf{n}=n-1$, where $n\in \mathbb{N}$ is the level
of EP \cite{5,forthcoming2,Marrani-Group32}. Thus, a direct relation between
EP and higher dimensional, exceptional $(1,0)$ SYM's in $D=\left(
9+8n\right) +$ $1$, $D=\left( 10+8n\right) +$ $2$, $D=\left( 11+8n\right) +$
$3$ and $D=\left( 12+8n\right) +$ $4$ exist\footnote{%
It should be pointed out that there exists an intrinsic, threefold
degeneracy in the determination of the noncompact real forms of EP
algebras. Indeed, at the $n$th level of EP, in the present paper (as it
will be evident from Section \ref{n}) we understand to enlarge by $8n$ only the
spacelike dimensions in the $\left( s+t\right) $-signature of the reductive,
pseudo-orthogonal part of the aforementioned EP algebras, thus obtaining $%
(9+8n)+1$, $(10+8n)+2$, $(11+8n)+3$ and $(12+8n)+4$, and matching (by
considering that $\mathbf{n}=n-1$) the space-time signatures (\ref{1-t-new}%
)-(\ref{4-t-new}). Nevertheless, the conjugation and reality properties of
spinors depend only on $D=s+t$ and on $\rho :=s-t$ (cf., e.g., \cite{SpinorAlgebras}), 
thus at the $n$th level of EP, the implementation of Bott
(i.e., mod($8n$)) periodicity could also be made by increasing by $4n$ both
spacelike and timelike dimensions, or also by increasing by $8n$ only the
timelike dimensions. In the former case one would obtain\ $(9+4n)+(1+4n)$, $%
(10+4n)+(2+4n)$, $(11+4n)+(3+4n)$ and $(12+4n)+(4+4n)$, thus giving rise (by
considering that $\mathbf{n}=n-1$) to the space-time signatures of (\ref{1-t}%
)-(\ref{4-t}), whereas in the latter case, one would obtain $9+(1+8n)$, $%
10+(2+8n)$, $11+(3+8n)$ and $12+(4+8n)$, thus giving rise (by considering
that $\mathbf{n}=n-1$) to other space-time signatures in which other chiral,
minimal $\mathcal{N}=(1,0)$ superalgebras, besides (\ref{1-t})-(\ref{4-t})
and (\ref{1-t-new})-(\ref{4-t-new}), exist. Such a threefold degeneracy of
the implementation of Bott periodicity (yielding spinors with the same
dimensions, reality properties and conjugation properties) can in principle
be applied at any level of EP, also in a different way from the way it was
implemented at the previous level; this allows to span a large variery of $%
\left( s,t\right) $-signatures in the $\mathbf{so}_{s,t}$ reductive part of
the noncompact real forms of EP algebras \cite{7}.}, hinting to the existence of an
``EP/SYM correspondence"~\cite{7}. In particular,
the levels $n=1,2,3$ (and for a generic $n$) of the class of SYM's in $%
D=\left( 11+8n\right) +3$ space-time dimensions with superalgebra (\ref%
{3-t-new}) are considered in the present paper.

\subsection{\label{n=2}The first nontrivial level ($n=2$)$:\mathfrak{e}%
_{8(-24)}^{(2)}$}

\subsubsection{$\mathcal{N}=(1,0)$ SYM in $19+3$ dimensions}

At the second level ($n=2$) of EP \cite{5}, one can define%
\begin{equation}
\mathfrak{e}_{8(-24)}^{(2)}:=\mathfrak{so}_{20,4}\oplus \mathbf{2048},
\label{a-2}
\end{equation}%
with algebraic structure and commutation relations defined in terms of Kac's
asymmetry function~\cite{5}, and with $\mathbf{2048}$ denoting the
MW semispinor in $20+4$ space-time dimensions. (\ref{a-2}) is the very first
step of a Bott-periodic, non-Lie generalization\footnote{%
This algebra was inspired by discussions with Eric Weinstein on extended
triality in $D=24$.} of $\mathfrak{e}_{8(-24)}\equiv \mathfrak{e}%
_{8(-24)}^{(1)}$, and it gives naturally rise to the following 5-grading,
\begin{equation}
\mathfrak{e}_{8(-24)}^{(2)}=\mathbf{22}_{-2}\oplus \mathbf{1024}%
_{-1}^{\prime }\oplus (\mathfrak{so}_{19,3}\oplus \mathbb{R})_{0}\oplus
\mathbf{1024}_{+1}\oplus \mathbf{22}_{+2},  \label{a-2bis}
\end{equation}%
where $\mathbf{1024}$ and $\mathbf{1024}^{\prime }$ are the MW semispinor
and its conjugate in $19+3$ space-time dimensions.

(\ref{a-2bis}) is the first nontrivial generalization of the 5-grading (\ref%
{a-3}) of $\mathfrak{e}_{8(-24)}$, and, in light of the discussion in Section %
\ref{Higher-SYMs}, it provides the vector and spinor structures for a novel,
exceptional $\mathcal{N}=(1,0)$ SYM in $19+3$, generalizing \cite{7}
the work of Bars, Sezgin, and Nishino \cite{2,1,1bis,6}. It is worth
stressing that $19+3$ is the signature of a unimodular lattice appearing in
the description of the space of periods for a complex $K3$ surface $S$ and a
Kahler class of $H^{1,1}(S,\mathbb{R})$ \cite{McMullen}.

Also, the moduli space of $\mathcal{N}=(4,4)$ string theories with K3 target
space has a discrete symmetry group that is the integral orthogonal group of
an even unimodular lattice of signature (20,4)\cite{aspinwallk3}. This may
permit further application of (\ref{a-2bis}) (with normalized cocycle) in
the study of vertex operator algebras for BPS states of K3 sigma models with
Mathieu group $\mathbb{M}_{24}$ symmetry \cite{Mathieu}.

\subsubsection{M2 brane in $18+1$ dimensions}

Considering the Lie group associated to the reductive (simple) part of $%
\mathfrak{e}_{8(-24)}^{(2)}$, namely $SO(20,4)$, we observe that a maximal
symmetric subgroup of this reads
\begin{equation}
SO(20,4)\supset SO(3,3)\times SO(17,1).
\end{equation}%
Again, $SO(3,3)$ yields affine transformations of $AdS_{4}$. On the other
hand, $SO(17,1)$ can be regarded as the affine symmetry of $S^{16}$, which
is the sphere acquired from $\mathbf{T}_{3}^{8,2}$, the rank-3 T-algebra of
special type \cite{Vinberg} that provides the first nontrivial
generalization of the Albert algebra $\mathbf{J}_{3}^{\mathbb{O}}\equiv
\mathbf{J}_{3}^{8}\equiv T_{3}^{8,1}$ within EP (cf. Fig. \ref%
{fig:MagicStarT}). Fixing a rank-1 idempotent of $\mathbf{T}_{3}^{8,2}$
induces the following $SO(17,1)$-covariant Peirce decomposition~\cite{5},
\begin{equation}
\mathbf{T}_{3}^{2}=\mathbf{18}\oplus \mathbf{256}\oplus \mathbf{1},
\label{aaa}
\end{equation}%
where $\mathbf{256}$ denotes the MW semispinor in signature $17+1$, and $%
\mathbf{1}$ is the fixed rank-1 idempotent of $\mathbf{T}_{3}^{8,2}$. (\ref%
{aaa}) can be regarded as a consequence of the maximal embedding \cite{5,
Marrani-Group32}%
\begin{equation}
T_{3}^{2}\supset \mathbb{R}\oplus \mathbf{\Gamma }_{17,1},
\end{equation}%
which in turn might give rise to a quasiconformal interpretation of the
definition (\ref{a-2}) itself \cite{forthcoming2}. The $18$-dimensional
Lorentzian spin factor $\mathbf{\Gamma }_{17,1}$ \cite{JVNW} has $SO(17,1)$
space-time symmetry, which is also the affine symmetry of $S^{16}$, a sphere
of the transverse degrees of freedom with a fixed (rank-1 idempotent) point
at infinity. It is here worth recalling that this structure is seen in the
3-grading of the first nontrivial extension $\mathfrak{e}_{6(-26)}^{(2)}$
of $\mathfrak{e}_{6(-26)}$ within EP \cite{Marrani-Group32},
\begin{equation}
\mathfrak{e}_{6(-26)}^{(2)}=\mathbf{256}_{-3}^{\prime }\oplus (\mathfrak{so}%
_{17,1}\oplus \mathbb{R})_{0}\oplus \mathbf{256}_{+3},
\end{equation}%
which might enjoy a reduced structure symmetry interpretation, as well \cite%
{forthcoming2}. Hence, $SO(20,4)$ contains maximally (and symmetrically) the
affine symmetries of $AdS_{4}\otimes S^{16}$.

Considering the reduction to transversal rotations\footnote{%
Note that the geometric picture in terms of projective lines in higher
projective spaces is lost for all nontrivial levels of EP, namely for $%
n\geqslant 2$, because octonionic (projective) geometry is defined only
until dimension two. This is also reflected in the fact that $\mathbf{T}%
_{3}^{8,2}$ is not a rank-3 Jordan algebra, and the triality among its block
components within $SO(16)$-covariant Peirce decomposition is spoiled \cite{5}%
. Again, another consequence of the aforementioned fact is that \textit{EP
algebras} are \textit{not} Lie algebras (because the Jacobi identity does
not hold on their non-reductive component \cite{5}). However, after
Rosenfeld \cite{Rosenfeld}, the isometry $SO(16)$ of $S^{15}$ can be
regarded as the stabilizer of $\left( \mathbb{O}\otimes \mathbb{O}\right)
\mathbb{P}^{2}=E_{8}/SO(16)$, which is an example (with the largest
exceptional global isometry) of the so-called Tits' buildings \cite{Tits,
Tits2}.} $S^{16}\rightarrow S^{15}$, and the isometry group $SO(17)$ reduces
to $SO(16)$. This corresponds to the two-step chain of maximal symmetric
embeddings into $SO(20,4)$,%
\begin{equation}
SO(20,4)\supset SO(19,3)\supset SO(3,3)\times SO(16).
\end{equation}%
Breaking down to isometries of $AdS_{4}$ by further maximal and symmetric
embedding yields,%
\begin{equation}
SO(20,4)\supset SO(19,3)\supset SO(3,3)\times SO(16)\supset SO(3,2)\times
SO(16).
\end{equation}%
Alternatively, the following embedding also holds,
\begin{equation}
SO(20,4)\supset SO(3,2)\times SO(16)\times SO(1,2),
\end{equation}%
where $SO(3,2)\times SO(16)$ acts as isometries of $AdS_{4}\otimes S^{15}$,
which can thus be regarded as a \textit{generalization} of the near-horizon
geometry of the M2 brane in $18+1$ space-time dimensions.

\subsubsection{M13 brane in $18+1$ dimensions}

Another maximal symmetric subgroup of $SO(20,4)$ is
\begin{equation}
SO(20,4)\supset SO(14,3)\times SO(6,1),
\end{equation}%
yielding that $SO(20,4)$ includes maximally (and symmetrically) the affine
transformations of $AdS_{15}\otimes S^{5}$. Restricting to transverse
directions induces $S^{5}\rightarrow S^{4}$, where the isometry group $%
SO(6)$ of $S^{5}$ breaks to $SO(5)$ of
%\footnote{%
%After Rosenfeld \cite{Rosenfeld}, the stabilizer $SO(12)$ of $S^{12}$ can be
%regarded as the stabilizer (up to a commuting $SU(2)$ factor) of the Tits'
%building $\left( \mathbb{H}\otimes \mathbb{O}\right) \mathbb{P}%
%^{2}=E_{7}/\left( SO(12)\times SU(2)\right) $ \cite{Tits, Tits2}.} 
$S^{4}$.
This corresponds to the two-step chain of maximal symmetric embeddings $%
SO(12,4)$,
\begin{equation}
SO(20,4)\supset SO(19,3)\supset SO(14,3)\times SO(5).
\end{equation}%
Breaking down to isometries of $AdS_{15}$ by further maximal and symmetric
embedding yields,%
\begin{equation}
SO(20,4)\supset SO(19,3)\supset SO(14,3)\times SO(5)\supset SO(14,2)\times
SO(5).
\end{equation}%
Alternatively, the following embedding also holds,
\begin{equation}
SO(20,4)\supset SO(14,2)\times SO(5)\times SO(1,2),
\end{equation}%
where $SO(14,2)\times SO(5)$ acts as isometries of $AdS_{15}\otimes S^{4}$,
which can thus be regarded as a \textit{generalization} of the near-horizon
geometry $AdS_{7}\otimes S^{4}$ [cf. (\ref{M5})] of the M5 brane in $10+1$
to the near-horizon geometry of an M13 brane, which is the Hodge dual
\footnote{%
An M13 brane is indeed recovered by taking the Hodge dual of the 4-form
field strength of the M2 in $D=18+1$ space-time, sourcing a 15-form dual
field strength.} of M2 in $18+1$ space-time dimensions.

\subsection{\label{n}The generic $n$th level ($n\in \mathbb{N}$)$:\mathfrak{%
e}_{8(-24)}^{(n)}$}

\subsubsection{Infinite $\mathcal{N}=(1,0)$ SYM's in $\left( 8n+3\right)+3$ dimensions}

The procedure of the previous subsection can be generalized to an arbitrary
level $n$ of EP (characterized by Bott periodicity) as follows \cite{5} ($%
n\in \mathbb{N}$; recall that $\mathfrak{e}_{8(-24)}\equiv \mathfrak{e}%
_{8(-24)}^{(1)}$),
\begin{eqnarray}
\mathfrak{e}_{8(-24)}^{(n)} &:&=\mathfrak{so}_{8n+4,4}\oplus \mathbf{2}%
^{(8n+6)/2}  \label{a-2-n-pre} \\
&=&\left( \mathbf{8n+6}\right) _{-2}\oplus \left( \mathbf{2}%
^{(8n+6)/2-1}\right) _{-1}^{\prime }\oplus (\mathfrak{so}_{8n+3,3}\oplus
\mathbb{R})_{0}\oplus \mathbf{2}_{-1}^{(8n+6)/2-1}\oplus \left( \mathbf{8n+6}%
\right) _{+2},  \label{a-2-n}
\end{eqnarray}%
where $\mathbf{2}^{(8n+6)/2}$ is the MW semispinor in $\left( 8n+4\right) +4$
space-time dimensions, while $\mathbf{2}^{(8n+6)/2-1}$ and $\left( \mathbf{2}%
^{(8n+6)/2-1}\right) ^{\prime }$ denote the MW spinor and its conjugate in $%
\left( 8n+3\right) +3$ space-time dimensions. (\ref{a-2-n-pre}) and (\ref%
{a-2-n}) respectively provide the $n$th generalization of (\ref{a-2}) and (%
\ref{a-2bis}) within EP \cite{5}. Again, in light of the discussion in Section %
\ref{Higher-SYMs}, this is tantalizing evidence for the possible existence
of a countably infinite tower (parametrized by $n\in \mathbb{N}$) of novel,
exceptional $\mathcal{N}=(1,0)$ SYM's in $\left( 8n+3\right) +3$
space-time dimensions. This generalization of the work by Bars, Sezgin, and
Nishino \cite{1,1bis,2,6}, briefly considered in Section \ref{Higher-SYMs}, is
the object of a forthcoming paper \cite{7}.

\subsubsection{M2 brane in $\left( 8n+2\right) +1$ dimensions}

Considering the Lie group associated to the reductive (simple) part of $%
\mathfrak{e}_{8(-24)}^{(n)}$, namely $SO(8n+4,4)$, we observe that a maximal
symmetric subgroup of this reads
\begin{equation}
SO(8n+4,4)\supset SO(3,3)\times SO(8n+1,1).
\end{equation}%
Once again, $SO(3,3)$ yields affine transformations of $AdS_{4}$. On the
other hand, $SO(8n+1,1)$ can be regarded as the affine symmetry of $S^{8n}$,
which is the sphere acquired from $\mathbf{T}_{3}^{8,n}$, the rank-3
T-algebra of special type \cite{Vinberg} that provides the $n$th
generalization of the Albert algebra $\mathbf{J}_{3}^{\mathbb{O}}\equiv
\mathbf{J}_{3}^{8}\equiv T_{3}^{8,1}$ within EP (cf. Fig. \ref%
{fig:MagicStarT}). Fixing a rank-1 idempotent of $\mathbf{T}_{3}^{8,n}$
induces the following $SO(8n+1,1)$-covariant Peirce decomposition \cite{5},
\begin{equation}
\mathbf{T}_{3}^{8,n}=\left( \mathbf{8n+2}\right) \oplus \mathbf{2}%
^{4n}\oplus \mathbf{1},  \label{aaaa}
\end{equation}%
where $\mathbf{2}^{4n}$ denotes the MW semispinor in signature $\left(
8n+1\right) +1$, and $\mathbf{1}$ is the fixed rank-1 idempotent of $\mathbf{%
T}_{3}^{8,n}$. (\ref{aaaa}) can be regarded as a consequence of the maximal
embedding \cite{5, Marrani-Group32}%
\begin{equation}
\mathbf{T}_{3}^{8,n}\supset \mathbb{R}\oplus \mathbf{\Gamma }_{8n+1,1},\label%
{embb}
\end{equation}%
which in turn might give rise to a quasiconformal interpretation of the
definition (\ref{a-2-n-pre}) itself \cite{forthcoming2}. The $\left(
8n+2\right) $-dimensional Lorentzian spin factor $\mathbf{\Gamma }_{8n+1,1}$
has $SO(8n+1,1)$ space-time symmetry, which is also the affine symmetry of $%
S^{8n}$, a sphere of the transverse directions with a fixed idempotent point
at infinity. It is here worth recalling that this structure is seen in the
3-grading of the $n$th generalization $\mathfrak{e}_{6(-26)}^{(n)}$ of $%
\mathfrak{e}_{6(-26)}$ within EP \cite{Marrani-Group32},%
\begin{equation}
\mathfrak{e}_{6(-26)}^{(n)}=\left( \mathbf{2}^{4n}\right) _{-3}^{\prime
}\oplus (\mathfrak{so}_{8n+1,1}\oplus \mathbb{R})_{0}\oplus \mathbf{2}%
_{+3}^{4n},
\end{equation}%
which might enjoy a reduced structure symmetry interpretation, as well \cite%
{forthcoming2}. Hence, $SO(8n+4,4)$ contains maximally (and symmetrically)
the affine symmetries of $AdS_{4}\otimes S^{8n}$.

Considering the reduction to transversal rotations\footnote{%
For $n\geqslant 3$, the interpretation of $S^{8n}$ (or its reduced space $%
S^{8n-1}$) in terms of higher projective projective spaces, or (of
stabilizers) of Tits' buildings, is generally lost. For further details,
cf. \cite{Marrani-Group32}.} $S^{8n}\rightarrow S^{8n-1}$, and the isometry
group $SO(8n+1)$ reduces to $SO(8n)$. This corresponds to the two-step chain
of maximal symmetric embeddings into $SO(8n+4,4)$,%
\begin{equation}
SO(8n+4,4)\supset SO(8n+3,3)\supset SO(3,3)\times SO(8n).
\end{equation}%
Breaking down to isometries of $AdS_{4}$ by further maximal and symmetric
embedding yields%
\begin{equation}
SO(8n+4,4)\supset SO(8n+3,3)\supset SO(3,3)\times SO(8n)\supset
SO(3,2)\times SO(8n).
\end{equation}%
Alternatively, the following embedding also holds,
\begin{equation}
SO(8n+4,4)\supset SO(3,2)\times SO(8n)\times SO(1,2),
\end{equation}%
where $SO(3,2)\times SO(8n)$ acts as isometries of $AdS_{4}\otimes S^{8n-1}$%
, which can thus be regarded as a \textit{generalization} of the near
horizon geometry of the M2 brane in $\left( 8n+2\right) +1$ space-time
dimensions, i.e., up to arbitrarily high Lorentzian space-times.

\subsubsection{M$(8n-3)$ brane in $\left( 8n+2\right) +1$ dimensions}

Another maximal symmetric subgroup of $SO(8n+4,4)$ is
\begin{equation}
SO(8n+4,4)\supset SO(8n-2,3)\times SO(6,1),  \label{eee}
\end{equation}%
yielding that $SO(8n+4,4)$ maximally (and symmetrically) includes the affine
transformations of $AdS_{8n-1}\otimes S^{5}$. Considering the reduction to
transverse directions 
%\footnote{%
%For $n\geqslant 3$, the interpretation of $S^{8n-3}$ (or of its reduction $%
%S^{8n-4}$) in terms of higher projective projective spaces, or (of
%stabilizers) of Tits' buildings, is generally lost. For further details, see
%\cite{Marrani-Group32}.} 
induces $S^{5}\rightarrow S^{4}$, where the
isometry group $SO(6)$ of $S^{5}$ breaks to $SO(5)$ of $S^{4}$. 
This corresponds to the two-step chain of maximal symmetric embeddings $%
SO(8n+4,4)$,
\begin{equation}
SO(8n+4,4)\supset SO(8n+3,3)\supset SO(8n-2,3)\times SO(5).
\end{equation}%
Breaking down to isometries of $AdS_{7}$ by further maximal and symmetric
embedding yields%
\begin{equation}
SO(8n+4,4)\supset SO(8n+3,3)\supset SO(8n-2,3)\times SO(5)\supset
SO(8n-2,2)\times SO(5).
\end{equation}%
Alternatively, the following embedding also holds,
\begin{equation}
SO(8n+4,4)\supset SO(8n-2,2)\times SO(5)\times SO(1,2),
\end{equation}%
where $SO(8n-2,2)\times SO(5)$ acts as isometries of $AdS_{8n-1}\otimes
S^{4} $, which can thus be regarded as a \textit{generalization} of the
near-horizon geometry of the M5 brane to an M$(8n-3)$ brane, which is the
Hodge dual of M2.\footnote{%
An M$(8n-3)$ brane is expected to source the $(8n-1)$-form field strength,
dual to the M2 4-form field strength in $D=\left( 8n+2\right) +1$ space-time.%
} It is worth noting that the magnetic (Hodge) dual branes of M2 branes
can be classified by quaternions as M$\left( 4k+1\right) $ branes: M5 brane
($k=1$), M13 brane ($k=3$), M21 brane ($k=5$), etc. over odd $k=2n-1$. This
provides a generalization of the observation of the projective line in $%
\mathbb{HP}^{2}$, done above (\ref{ee}), in $\left( 8n+2\right) +1$
space-time dimensions, i.e., up to arbitrarily high Lorentzian
space-times.

\subsection{\label{n=3}The third level ($n=3$)$:\mathfrak{e}_{8(-24)}^{(3)}$,
bosonic M-Theory, and monster AdS/CFT}

\subsubsection{$\mathcal{N}=(1,0)$ SYM in $27+3$ dimensions}

As an interesting example, let us consider the generalization of $\mathfrak{e%
}_{8(-24)}$ provided by the $n=3$ level of EP \cite{5}, namely
\begin{eqnarray}
\mathfrak{e}_{8(-24)}^{(3)} &:=&\mathfrak{so}_{28,4}\oplus \mathbf{2}^{15}
\label{a-2-3-pre} \\
&=&\mathbf{30}_{-2}\oplus \left( \mathbf{2}^{14}\right) _{-1}^{\prime
}\oplus (\mathfrak{so}_{27,3}\oplus \mathbb{R})_{0}\oplus \mathbf{2}%
_{+1}^{14}\oplus \mathbf{30}_{+2},  \label{a-2-3}
\end{eqnarray}%
where $\mathbf{2}^{15}$ is the MW semispinor in $28+4$, while $\mathbf{2}%
^{14}$ and $\left( \mathbf{2}^{14}\right) ^{\prime }$ denote the MW spinor
and its conjugate in $27+3$. (\ref{a-2-3-pre}) and (\ref{a-2-3}),
respectively, are the $n=3$ elements of the Bott-periodized
countably infinite sequences of generalizations (\ref{a-2-n-pre}) and (\ref{a-2-n})
within EP. As discussed in Section \ref{Higher-SYMs}, \ref{a-2-3-pre}) and (\ref%
{a-2-3}) intriguingly provide evidence for the possible existence of a novel
$\mathcal{N}=(1,0)$ SYM in $27+3$ \cite{7}, generalizing previous works of
Bars, Sezgin, and Nishino \cite{2,1,1bis,6}.

\subsubsection{M2 brane in bosonic M-theory}

Considering the Lie group associated to the reductive (simple) part of $%
\mathfrak{e}_{8(-24)}^{(3)}$, namely $SO(28,4)$, we observe that a maximal
symmetric subgroup of this reads
\begin{equation}
SO(28,4)\supset SO(3,3)\times SO(25,1),
\end{equation}%
where $SO(3,3)$ yields affine transformations of $AdS_{4}$ and $SO(25,1)$
can be regarded as the affine symmetry of $S^{24}$, or alternatively as the
space-time symmetry of bosonic string theory \cite{14}. By specializing to $%
n=3$ the previous $n$-parametrized treatment, $SO(25,1)$ can be regarded as
the affine symmetry of $S^{24}$, which is the sphere acquired from $\mathbf{T%
}_{3}^{8,3}$, the rank-3 T-algebra of special type \cite{Vinberg} that
provides the third generalization of the Albert algebra within EP (cf. Fig. %
\ref{fig:MagicStarT}). Fixing a rank-1 idempotent of $\mathbf{T}_{3}^{8,3}$
induces the following $SO(25,1)$-covariant Peirce decomposition \cite{5},
\begin{equation}
\mathbf{T}_{3}^{8,3}=\mathbf{26}\oplus \mathbf{2}^{12}\oplus \mathbf{1},
\label{aaa-3}
\end{equation}%
where $\mathbf{2}^{12}$ denotes the MW semispinor in signature $25+1$, and $%
\mathbf{1}$ is the fixed rank-1 idempotent of $\mathbf{T}_{3}^{8,3}$. (\ref%
{aaa-3}) can be regarded as a consequence of the maximal embedding \cite{5,
Marrani-Group32}%
\begin{equation}
\mathbf{T}_{3}^{8,3}\supset \mathbb{R}\oplus \mathbf{\Gamma }_{25,1}.
\end{equation}%
The $26$-dimensional Lorentzian spin factor $\mathbf{\Gamma }_{25,1}$ has
bosonic string theory space-time symmetry, which is also the affine symmetry
of $S^{24}$, a sphere of the transverse degrees of freedom with fixed
idempotent point at infinity. The span of this fixed idempotent can serve as
a 27th dimension for $D=27$ M-theory \cite{14}. Note, bosonic string theory
space-time symmetry is seen in the 3-grading of the third generalization $%
\mathfrak{e}_{6(-26)}^{(3)}$ of $\mathfrak{e}_{6(-26)}$ within EP \cite%
{Marrani-Group32},
\begin{equation}
\mathfrak{e}_{6(-26)}^{(3)}=\left( \mathbf{2}^{12}\right) _{-3}^{\prime
}\oplus (\mathfrak{so}_{25,1}\oplus \mathbb{R})_{0}\oplus \mathbf{2}%
_{+3}^{12}.
\end{equation}%
Hence, $SO(28,4)$ contains maximally (and symmetrically) the affine
symmetries of $AdS_{4}\otimes S^{24}$.

Considering the reduction to transversal rotations, $S^{24}\rightarrow S^{23}
$, and the isometry group $SO(25)$ reduces to $SO(24)$. This corresponds to
the two-step chain of maximal symmetric embeddings into $SO(28,4)$,%
\begin{equation}
SO(28,4)\supset SO(27,3)\supset SO(3,3)\times SO(24).
\end{equation}%
Breaking down to isometries of $AdS_{4}$ by further maximal and symmetric
embedding yields%
\begin{equation}
SO(28,4)\supset SO(27,3)\supset SO(3,3)\times SO(24)\supset SO(3,2)\times
SO(24).
\end{equation}%
Alternatively, the following embedding also holds,
\begin{equation}
SO(28,4)\supset SO(3,2)\times SO(24)\times SO(1,2),
\end{equation}%
where $SO(3,2)\times SO(24)$ acts as isometries of $AdS_{4}\otimes S^{23}$,
which can thus be regarded as a \textit{generalization} of the near-horizon
geometry of the M2 brane in $26+1$ space-time dimensions. $AdS_{4}\otimes
S^{23}$ might also serve as a possible vacuum of bosonic M-theory \cite{14}.

\subsubsection{M21 brane in bosonic M-theory}

Another maximal symmetric subgroup of $SO(28,4)$ is
\begin{equation}
SO(28,4)\supset SO(22,3)\times SO(6,1),
\end{equation}%
yielding that $SO(28,4)$ maximally (and symmetrically) includes the affine
transformations of $AdS_{23}\otimes S^{5}$. Considering the reduction to
transverse directions maps $S^{5}\rightarrow S^{4}$, and the isometry
group $SO(6)$ of $S^{5}$ breaks to $SO(5)$ on $S^{4}$. This
corresponds to the two-step chain of maximal symmetric embeddings $SO(28,4)$,
\begin{equation}
SO(28,4)\supset SO(27,3)\supset SO(22,3)\times SO(5).
\end{equation}%
Breaking down to isometries of $AdS_{23}$ by further maximal and symmetric
embedding yields%
\begin{equation}
SO(28,4)\supset SO(27,3)\supset SO(22,3)\times SO(5)\supset SO(22,2)\times
SO(5).
\end{equation}%
Alternatively, the following embedding also holds,
\begin{equation}
SO(28,4)\supset SO(22,2)\times SO(5)\times SO(1,2),
\end{equation}%
where $SO(22,2)\times SO(5)$ acts as isometries of $AdS_{23}\otimes S^{4}$,
which can thus be regarded as a \textit{generalization} of the near-horizon
geometry of the M5 brane to an M21 brane, the Hodge dual of the M2 brane in $%
26+1$ space-time dimensions. In view of the considerations above and
Horowitz and Susskind's conjectured M21 brane in \cite{14}, $AdS_{23}\otimes
S^{4}$ would also provide support for a possible vacuum of bosonic
M-theory, \textquotedblleft dual" to $AdS_{4}\otimes S^{23}$.

% Should have looked at duals, but we just did M2 and M5's all the way up, only dual for D=11.
% Briefly remark on dual of M2 brane in 27D...
\iffalse{
Note also that $SO(28,4)$ contains the isometries of $AdS_{7}$ and of the
M21 brane worldvolume, times a dilatational factor,%
\begin{equation}
SO(28,4)\supset SO(6,2)\times SO(21,1)\times SO(1,1),
\end{equation}%
or, equivalently, through two different maximal and symmetric embeddings,
the isometries of $AdS_{7}$ times the conformal symmetry of the M21 brane
worldvolume, or, respectively, the conformal symmetry of $AdS_{7}$ times the
isometries of the M21 brane worldvolume,%
\begin{eqnarray}
SO(28,4) &\supset &SO(6,2)\times SO(22,2), \\
SO(28,4) &\supset &SO(7,3)\times SO(21,1).
\end{eqnarray}
\fi

\subsubsection{Conway group and Witten's monster AdS/CFT}

By reducing to isometries of%
\begin{equation}
AdS_{3}=O(2,2)/O(2,1),
\end{equation}%
the following embedding into $SO(28,4)$ is singled out,
\begin{equation}
SO(28,4)\supset SO(2,2)\times SO(24)\times SO(2,2),
\end{equation}%
where $SO(2,2)\times SO(24)$ yields isometries of $AdS_{3}\otimes S^{23}$.

We note that the space $AdS_{3}\otimes S^{23}$ is especially interesting, as
it lends support for the monster $AdS$/CFT for three-dimensional gravity
proposed by Witten \cite{13}. If we suppose $AdS_{3}\otimes S^{23}$ is a
possible vacuum of bosonic string theory\footnote{%
Thanks to Lubos Motl for comments on this construction.}, where the expected $%
\mathcal{R}$-symmetry is $SO(24)$ (from $D=26+1$ M-theory reduced to $D=2+1$), we can identify the Conway group $%
Co_{0}\subset \mathbb{M}$ (where $\mathbb{M}$ is the monster group) in $%
SO(24)$ as acting on a discretized $S^{23}$ with points given by vectors of
the Leech lattice \cite{15,16}, in which the first shell of $196,560$
vectors has norm four \cite{16}. This allows a finite group action as the
Conway group $Co_{0}$ is the automorphism group of the Leech lattice \cite%
{16}, acting as isometries of the discretized\footnote{%
It is tempting to observe that the spherical parts of the corresponding
geometries for the levels $n=1$ and $n=2$ of EP, namely of $AdS_{3}\otimes
S^{7}$ and $AdS_{3}\otimes S^{15}$, can be discretized by using $E_{8}$ and $%
E_{8}\oplus E_{8}$ lattices, respectively.} $S^{23}$. The appearance of the
Conway group is more than fortuitous, as every $K3$ sigma model has finite
group symmetry contained in $Co_{0}$\cite{Mathieu}. This is a tantalizing
hint that, beyond the search for vertex operator algebras on BPS states with
exact $\mathbb{M}_{24}$ symmetry \cite{Mathieu}, one can move to a larger
vertex operator algebra (CFT) with manifest Conway group symmetry and study
its related $K3$ sigma models. Intriguingly, a super vertex operator algebra
with $Co_{0}$ Conway group symmetry has already been constructed in \cite%
{conwayVOA}. An $AdS$/CFT study of $AdS_{3}\otimes S^{23}$ might involve
this particular vertex operator algebra. In light of this algebra, the
monster group could be a finite symmetry of the light cone little group of
nonperturbative\footnote{%
Formally, such a formulation would exist at levels $n=5$ and $n=6$ of EP,
where a 24-dimensional even unimodular lattice and its extensions can be
constructed.} $27$-dimensional M-theory.

\section{\label{Conclusion}Conclusion}

By relying on the maximal embedding (\ref{embb}) of semi-simple rank-3
Jordan algebras into rank-3 T-algebras of special class, we used 3- and
5-gradings of finite-dimensional exceptional Lie algebras and their
Bott-periodic extensions within EP, in order to show that the near-horizon
geometries of the M2 branes and their Hodge (magnetic) duals can be
generalized to arbitrarily high dimensions.

Moreover, we showed how the EP generalizations of the minimally noncompact,
real form $\mathfrak{e}_{8(-24)}$ of the largest finite-dimensional
exceptional Lie algebra $\mathfrak{e}_{8}$ hint at novel, exceptional 
SYM's beyond $11+3$ space-time dimensions, with tantalizing signatures $%
17+1$, $19+3$, $25+1$ and $27+3$, suggesting a periodic ladder to the $%
D=25+1 $ bosonic string (with an $AdS_{3}\otimes S^{23}$ vacuum) and $D=26+1$
bosonic M-theory (with M2 $AdS_{4}\otimes S^{23}$ near horizon geometry),
that can be generalized to arbitrarily high dimension with $1$, $2$, $3$ or $%
4$ timelike dimensions. This allowed us to argue that an 
``EP/SYM correspondence" can be put forward \cite{7},
implying an infinite dimensional spectral extension of M-theory via cubic
matrix T-algebras, where $D=s+t=10+1$ is the maximal single-time extension
of SYM exhibiting manifest ($SO(8)$) triality.

The appearance of $19+3$ and $20+4$ signatures in moduli spaces with $K3$
target space \cite{McMullen, aspinwallk3}, and Mathieu moonshine in $K3$
sigma models \cite{Mathieu}, with $AdS$ constructions descending from $28+4$
exhibiting Conway group $Co_{0}$ symmetry, all yield evidence that
higher dimensional SYM's in the EP context (whose existence has been briefly
discussed in Section \ref{Higher-SYMs} and will be investigated in detail in
\cite{7}) can help shed light on the \textquotedblleft mysterious duality"
\cite{8} of Iqbal, Nietzke and Vafa, as well as on the moonshine structure
of M-theory and beyond.

Forthcoming companion papers and studies will explore these issues in deeper
detail \cite{7,forthcoming2,Marrani-Group32}.

\section*{Acknowledgements}
\begin{center}
Presented by M.R. at the Symposium \textit{Symmetries and Order: Algebraic
Methods in Many Body Systems}

in honor of Prof. Francesco Iachello,

Wright Laboratory and Yale University Physics Department, CT, USA, October
5-6, 2018
\end{center}

%\newpage

\end{document}